\documentclass[aps,prb,showpacs,twocolumn,floats,psfig]{revtex4}
\usepackage{amssymb}
\usepackage{amsbsy}
\usepackage{amsmath}
\usepackage{epsfig}

\newcommand\bea{\begin{eqnarray}}
\newcommand\eea{\end{eqnarray}}
\newcommand\beq{\begin{equation}}
\newcommand\eeq{\end{equation}}

\newcommand{\noi}{\noindent}
\newcommand{\non}{\nonumber}
\newcommand{\bib}{\bibitem}
\newcommand{\al}{\alpha}
\newcommand{\de}{\delta}
\newcommand{\lam}{\lambda}

\newcommand{\da}{\dagger}
\newcommand{\la}{\langle}
\newcommand{\ra}{\rangle}
\newcommand{\vk}{\vec k}
\newcommand{\vn}{\vec n}
\newcommand{\vcr}{\vec r}

\begin{document}
\title{Theory of defect production in nonlinear quench across a quantum
critical point}
\author{Shreyoshi Mondal$^1$, K. Sengupta$^1$, and Diptiman Sen$^2$}
\affiliation{$^1$T.C.M.P. Division, Saha Institute of Nuclear Physics, 1/AF
Bidhannagar, Kolkata 700 064, India\\ $^2$Center for High Energy Physics,
Indian Institute of Science, Bangalore, 560 012, India}

\date{\today}

\begin{abstract}
We study defect production in a quantum system subjected to a
nonlinear power law quench which takes it either through a quantum
critical or multicritical point or along a quantum critical line. We
elaborate on our earlier work [D. Sen, K. Sengupta, S. Mondal, \prl
101, 016806 (2008)] and present a detailed analysis of the scaling
of the defect density $n$ with the quench rate $\tau$ and exponent
$\al$ for each of the above-mentioned cases. We also compute the
correlation functions for defects generated in nonlinear quenches
through a quantum critical point and discuss the dependence of the
amplitudes of such correlation functions on the exponent $\al$. We
discuss several experimental systems where these theoretical
predictions can be tested.
\end{abstract}


\maketitle

\section{Introduction}
\label{intro}

Quantum phase transitions have been widely studied in different
systems for several years \cite{sachdev1}. Such transitions occur
when the ground state of a quantum system changes due to the variation
of some system parameter such as pressure \cite{hfref}, doping
\cite{hightcref} or magnetic field \cite{spinref}. More recently,
non-equilibrium physics around such critical points has also been
studied \cite{sachdev2,ks1}. In particular, quench dynamics through
quantum critical points has been a subject of intense theoretical
study in recent years. Such a dynamics involves the time evolution of a
parameter $\lam \equiv \lam(t)$ in the Hamiltonian of the system
which carries it through a quantum critical point, characterized by
the correlation length exponent $\nu$ and the dynamical critical
exponent $z$, at $\lam=\lam_c$. Since the energy gaps between the
ground and the first excited states vanish at the
quantum critical point, the dynamics of the system necessarily
becomes non-adiabatic in a finite region around this point even for an
arbitrarily slow quench. This leads to the failure of the system to
follow the instantaneous ground state. As a result defects are
produced \cite{kibble,zurek,damski}. Most of the initial studies
of defect production in quench dynamics for various quantum systems
have been restricted to the case of a linear quench $\lam(t)\equiv \lam_0
t/\tau$, where $\tau^{-1}$ is the quench rate \cite{ks1,dziar1,cardy1,das1,
levitov1,pell1,sen1,bd2,caneva1,wz1,sen2,sen3,sen4,santoro1}.
It is well known that for a slow linear quench, the defect density
$n\sim \tau^{-d\nu /(z\nu +1)}$ where $d$ is the dimension of the
system \cite{anatoly1,anatoly2}. More recently, nonlinear power law
quenches characterized by $\lam(t)=\lam_0 |t/\tau|^{\al} {\rm
sign}(t)$, where $\al$ denotes the power law exponent and ${\rm sign}$
is the signum function, have also been studied \cite{ks3,anatoly3}. In
particular, it has been shown in Ref. \onlinecite{ks3}, that if, during the
quench, the critical point is reached at time $t=0$ [$t=t_0 \ne 0$], then the
defect density $n$ for such a quench process scales as $n\sim \tau^{-d\al
\nu/(\al z \nu +1)}$ $[n\sim \left(\al g^{(\al -1)/\al}/ \tau \right)^{\nu
d/(z\nu +1)}$, where $g$ is a non-universal constant.]

On the experimental side, trapped ultracold atoms in optical
lattices have provided ways to realize many interacting quantum
systems with a variety of low temperature phases separated by
quantum critical points \cite{bloch1,duan}. These systems provide an
easy access to non-equilibrium dynamics of its constituent atoms and
hence provide ideal experimental test beds for quench related studies.
Defect production has already been studied experimentally for a spin-1
Bose condensate \cite{sk1}. However, a detailed experimental study of
nonlinear quench dynamics has not been undertaken so far.

In this paper we study defect production due to nonlinear power law
quenches in quantum critical systems. Our main results are the
following. First, we elaborate on the work of Ref. \onlinecite{ks3}
and provide a detailed derivation of the scaling laws of the defect
densities mentioned above. Second, we extend the scaling law for
defect production through multicritical points, as studied for
a linear quench in Ref. \onlinecite{sen4}, to nonlinear quenches.
Third, motivated by the work in Ref. \onlinecite{sen3}, we derive
scaling laws for defect densities produced during a nonlinear quench
when the system is taken along a {\it gapless line} during the
quench. Our results extend those in Ref. \onlinecite{sen4} and
\onlinecite{sen3}, and reproduce them as special cases. Fourth,
taking the one-dimensional Kitaev model as a specific system, we
compute the correlation functions for defects produced during a
nonlinear quench. We also provide a general model independent
discussion of the behavior of such correlation functions. Finally,
we present a detailed discussion of possible experimental systems
where these theoretical results may be tested.

The organization of the paper is as follows. In Sec. \ref{expo} we
provide detailed derivations for the scaling laws of defect density
produced during a nonlinear quench. This is followed, in Sec.
\ref{c1}, by a computation of the defect correlation functions.
Next, in Sec. \ref{num1}, we provide numerical studies to corroborate
our analytical results. In Sec. \ref{expts}, we discuss possible
experimental systems where the scaling laws derived in Sec. \ref{expo}
can be tested. Finally we conclude in Sec. \ref{conclusion}.

\section{Defect production rate in a nonlinear quench}
\label{expo}

The density of defects produced in a quench process depends
crucially on the nature of the phases that the system passes through
during the quench. Such processes can therefore be broadly
classified into three types. First, the system may pass from one
gapped phase to another through an intermediate gapless critical or
multicritical point. Second, the system may move along a gapless
critical line in the parameter space so that at each point on that
line the gap vanishes at a fixed and unique momentum
\cite{pell1,sen3}. Third, the quench may take the system from a
gapped phase to another through a gapless hypersurface in parameter
space as well as in momentum space \cite{sen2}. In what follows, we
will study defect production during nonlinear quench of the first
two types in Secs. \ref{hamil1}, \ref{hamil2} and \ref{hamil3}. An
analogous study for the third case, where the system passes through
a hypersurface in momentum space, is beyond the scope of the present
work.

\subsection{Quench dynamics from one gapped phase to another}
\label{hamil1}

We start with the model Hamiltonian for a $d$-dimensional system
\bea H(t) &=& \sum_{\vk} ~\psi_{\vk}^{\da} ~H_{\vk} (t) ~\psi_{\vk}, \non \\
H_{\vk}(t) &=& (\lam(t)+b(\vk))\tau_3+\Delta(\vk)\tau_+
+\Delta^*(\vk)\tau_- , \label{ham1} \eea where $b(\vk)$ and
$\Delta(\vk)$ are model-dependent functions, $\tau_i$ denote usual
Pauli matrices, $\lam(t)= \lam_0 |t/\tau|^{\al} {\rm sign}(t)$ is
the quench parameter where $\al=1$ implies linear quench, and
$\psi_{\vk}=(c_{1\vk}, c_{2\vk})$ represents the fermionic
operators. Such a Hamiltonian is known to represent several one- and
two-dimensional spin models such as the Ising \cite{sachdev1}, the
$XY$ \cite{sen1} and the extended Kitaev model
\cite{chen1,lee1,sen2}. The instantaneous energies of the
Hamiltonian given by Eq. (\ref{ham1}) are given by \bea E(\vk) &=&
\pm ~\sqrt{(\lam(t)+b(\vk))^2 + |\Delta(\vk)|^2}. \label{en1} \eea
These energy levels touch each other at $t=t_0$ and $\vk=\vk_0$, so
that $|\Delta(\vk)| \sim |\vk - \vk_0|$ and $|t_0| = \tau
|b(\vk_0)/\lam_0|^{1/\al} = \tau g^{1/\al}$, where $g=
|b(\vk_0)/\lam_0|$ is a non-universal model-dependent parameter. At
this point the energy levels cross and we have a quantum critical
point with $\nu=z=1$. Note that the critical point is reached at
$t=0$ only if $b(\vk_0)$ vanishes.

Let us first consider the case where $b(\vk_0)=0$ so that the system
passes through the critical point at $t=0$. In what follows, we
shall assume that $|\Delta(\vk)| \sim |\vk -\vk_0|$ and $ b(\vk)
\sim |\vk-\vk_0|^{z_1}$ at the critical point, where $z_1 \ge 1$ so
that $E \sim |\vk-\vk_0|$ and $z=1$. In the rest of the analysis, we
set $\hbar=1$, and scale $t \to t\lam_0$, $\tau \to \tau \lam_0$,
$\Delta(\vk) \to \Delta(\vk)/\lam_0$, and $b(\vk) \to
b(\vk)/\lam_0$.

We begin by observing that the ground state of the system must be
$(c_{1 \vk},c_{2\vk})=(1,0)\,[(0,1)]$ at the beginning [end] of the
quench at $t =-\infty\,[\infty]$. Thus the probability of defect
formation, ${\it i.e}$, the probability for the system to be in the
excited state at the end of the quench for a given state
$|\vk\rangle$ must be given by \bea p_{\vk} &=& \lim_{t \to \infty}
|c_{1\vk} (t)|^2 . \eea The density of these defects is thus given
by \beq n={\rm{\lim_{t \to \infty}}}\int_{\rm
BZ}\frac{d^dk}{(2\pi)^{d}}~ |c_{1\vk} (t)|^2 , \eeq where $\int_{\rm
BZ}$ denotes integration over the Brillouin zone.

To obtain $p_{\vk}$, we study the time evolution of the system which
is governed by the Schr\"odinger equation $i\partial \psi_{\vk}
/\partial t =H_{\vk} \psi_{\vk}$; this leads to the following
equations, \bea i\dot{c}_{1\vk} &=& (|t/\tau|^{\al} {\rm sign}(t)
+b(\vk))c_{1\vk}+
\Delta(\vk)c_{2\vk}, \non \\
i\dot{c}_{2 \vk}&=&-(|t/\tau|^{\al} {\rm sign}(t) +b(\vk))c_{2\vk}+
\Delta^*(\vk)c_{1\vk}, \non \\
& & \label{c2} \eea where we have kept the time dependence of
$c_{1\vk(2 \vk)}(t)$ implicit, and $\dot{c}_{1\vk(2 \vk)}(\vk)
\equiv
\partial_t c_{1\vk(2\vk)}$. To solve these equations, we define \bea
c_{1\vk}^{'} &=& c_{1\vk} ~e^{i\int^t dt^{'}(|t^{'}/ \tau|^\al {\rm
sign}
(t^{'}) + b(\vk))} \non \\
c_{2\vk}^{'} &=& c_{2 \vk} ~e^{-i\int^t dt^{'}(|t^{'}/\tau|^\al {\rm
sign} (t^{'}) + b(\vk))}. \label{c2'} \eea Then substituting Eq.
(\ref{c2'}) in Eq. (\ref{c2}) and eliminating $c_{2\vk}^{'}$ from
the resulting equations, we get \bea \ddot{c}_{1\vk}^{'}&-& 2i
~[|t/\tau|^{\al}{\rm sign}(t)+b(\vk)] ~
\dot{c}_{1\vk}^{'} \non \\
&+& |\Delta(\vk)|^2 ~c_{1\vk}^{'} ~=~ 0. \label{ddotc1} \eea Now we
scale $t \to t\tau^{\al/(\al+1)}$ so that Eq. (\ref{ddotc1}) becomes
\bea \ddot{c}_{1\vk}^{'} &-& 2i ~[|t|^\al {\rm sign}(t) +
b(\vk)\tau^{\al/
(\al +1)}] ~\dot{c}_{1\vk}^{'} \non \\
&+& |\Delta(\vk)|^2\tau^{2\al/(\al +1)} ~c_{1\vk}^{'} ~=~ 0.
\label{ddotc11} \eea {}From Eq. (\ref{ddotc11}) we immediately note
that since $c_{1\vk}$ and $c_{1\vk}^{'}$ differ only by a phase
factor, $p_{\vk}$ must be given by \bea p_{\vk} &=& {\lim}_{t \to
\infty}|c'_{1\vk} (t)|^2 =
f[b(\vk)\tau^{\frac{\al}{\al+1}},|\Delta(\vk)|^2\tau^{\frac{2\al}{
\al+1}} ], \label{pexp} \eea where $f$ is a function whose
analytical form is not known for $\al \ne 1$. Nevertheless, we note
that for a slow quench (large $\tau$), $p_{\vk}$ becomes appreciable
only when the instantaneous energy gap, as obtained from Eq.
(\ref{en1}), becomes small at some point of time during the quench.
Consequently, $f$ must vanish when either of its arguments are
large: $f(\infty,a)=f(a,\infty)=0$ for any value of $a$. Thus for a
slow quench (large $\tau$), the defect density $n$ is given by \bea
n &=& \int_{\rm BZ}\frac{d^dk}{(2\pi)^{d}} ~f
[b(\vk)\tau^{\frac{\al}{
\al+1}},|\Delta(\vk)|^2\tau^{\frac{2\al}{\al+1}} ], \eea and
receives its main contribution from values of $f$ near $\vk=\vk_0$
where both $b(\vk)$ and $\Delta(\vk)$ vanish. Thus one obtains,
after extending the range of momentum integration to $\infty$, \bea
n \simeq \int \frac{d^d k}{(2\pi)^d} \, f\left[|\vk-\vk_0|^{z_1}
\tau^{\frac{\al}{\al+1}};|\vk-\vk_0|^{2}
\tau^{\frac{2\al}{\al+1}}\right]. \eea Now scaling $\vk \to (\vk -
\vk_0) \tau^{\al/(\al+1)}$, we find that \bea n &=& \tau^{- \frac{d
\al}{\al+1}}\int \frac{d^d k}{(2\pi)^d}~
f(|\vk|^{z_1} \tau^{\al(1-z_1)/(\al+1)};|\vk|^2) \non \\
& \simeq & \tau^{- \frac{d \al}{\al+1}} \int \frac{d^d k}{(2\pi)^d}~
f(0;|\vk|^2)~ \sim ~\tau^{- \frac{d \al}{\al+1}}, \label{sca1} \eea
where in arriving at the last line, we have used $z_1 > 1$ and $\tau
\to \infty$. (If $z_1 = 1$, the integral in the first line is
independent of $\tau$, so the scaling argument still holds). Note
that for $\al=1$, Eq. (\ref{sca1}) reduces to its counterpart for a
linear quench \cite{anatoly1}. It turns out that the case $z_1 < 1$
deserves a detailed discussion which we defer till Sec.
\ref{hamil2}.

Next we generalize our results for a critical point with arbitrary
values of $\nu$ and $z$. To this end, we consider a generic
time-dependent Hamiltonian $H_1[t]\equiv H_1[\lam(t)]$, whose states
are labeled by $|\vk \ra$ and $|0\ra$ denotes the ground state. If
there is a second order phase transition, the basis states change
continuously with time during this evolution and can be written as
\bea |\psi(t)\ra ~=~ \sum_{\vk} ~a_{\vk}(t) ~|\vk[\lam(t)]\ra. \eea
The defect density can then be obtained in terms of these
coefficients $a_{\vk} (t)$ as \bea n ~=~ \sum_{\vk \ne 0}
~|a_{\vk}(t\to \infty)|^2 . \eea Following the analysis in Ref.
\onlinecite{anatoly1}, one can then obtain an expression for the
defect density $n$ as \bea n \simeq \int \frac{d^d k}{(2 \pi)^d}
\Big| \int_{-\infty}^\infty d \lam \la \vk|\frac{d}{d \lam} |0 \ra
e^{i \tau \int^\lam d \lam' \de E_{\vk} (\lam')} \Big|^2,
\label{defect2} \eea where $\de E_{\vk} (\lam)=E_{\vk}
(\lam)-E_0(\lam)$ are the instantaneous excitation energies, and we
have replaced the sum over $\vk$ by a $d$-dimensional momentum
integral. We note, following Ref. \onlinecite{anatoly1}, that near a
critical point, \bea \de E_{\vk} (\lam) ~=~ \Delta
F(\Delta/|\vk-\vk_0|^z), \label{rel1} \eea where $\Delta$ is the
energy gap, $z$ is the dynamical critical exponent, and $F(x) \sim
1/x$ for large $x$. Also, since the quench term vanishes at the
critical point, $\Delta \sim |\lam|^{\al z \nu}$ for a nonlinear
quench, one can write \bea \de E_{\vk} (\lam) ~=~ |\lam|^{\al z \nu}
F'(|\lam|^{\al z \nu}/|\vk- \vk_0|^z), \label{rel2} \eea where
$F'(x) \sim 1/x$ for large $x$. Further, one has $\la \vk|\frac{d}{d
\Delta} |0\ra = |\vk-\vk_0|^{-z} G(\Delta /|\vk-\vk_0|^z)$ near a
critical point, where $G(0)$ is a constant. This allows us to write
\bea \la \vk|\frac{d}{d \lam}|0\ra ~=~ \frac{\lam^{\al z \nu
-1}}{|\vk- \vk_0|^z} G'(\lam^{\al z \nu}/|\vk-\vk_0|^z),
\label{rel3} \eea where $G'(0)$ is a constant
\cite{sachdev1,anatoly1}. Substituting Eqs.\ (\ref{rel2}) and
(\ref{rel3}) in Eq. (\ref{defect2}) and changing the integration
variables to $\eta = \tau^{\al \nu/(\al z \nu + 1)} |\vk-\vk_0|$ and
$ \xi = |\vk-\vk_0|^{-1/(\al \nu)} \lam$, we find that \bea n
~\simeq ~C ~\tau^{-\al \nu d/(\al z \nu +1)}, \label{defect3} \eea
where $C$ is a non-universal number independent of $\tau$.

Next we focus on the case where the quench term does not vanish at
the quantum critical point for $\vk = \vk_0$. We again consider the
Hamiltonian $H_{\vk}(t)$ in Eq. (\ref{ham1}), but now assume that the
critical point is reached at $t=t_0 \ne 0$. This renders our
previous scaling argument invalid since $\Delta (\vk_0) = 0$ but
$b(\vk_0) \ne 0$. In this situation, $|t_0/\tau| = g^{1/\al}$ so
that the energy gap $\de E$ may vanish at the critical point for
$\vk = \vk_0$. We now note that the most important contribution to
the defect production comes from times near $t_0$ and from momenta
near $k_0$. Hence we expand the diagonal terms in $H_{\vk}(t)$
about $t=t_0$ and $\vk =\vk_0$ to obtain
\bea H'(t) &=& \sum_{\vk} \psi^{\da} (\vk) ~\Big[\left\{\al g^{(\al-1)/\al}
\left(\frac{t-t_0}{\tau}\right)+ b' (\de \vk)\right\} \tau_3 \non \\
&& ~~~~~~~~~~~~~~ +\Delta(\vk) \tau_+ + \Delta^{\ast}(\vk) \tau_-
\Big] ~\psi(\vk), \label{ham2} \eea
where $b'(\de \vk)$ represents all the terms in the expansion of $b(\vk)$
about $\vk=\vk_0$, and we have neglected all terms
\bea R_n &=& (\al-n+1)(\al-n+2)...(\al) \non \\
&& \times ~g^{(\al-n)/\al} |(t-t_0)/\tau|^n {\rm sign}(t)/n!
\label{negterm1} \eea
for $n>1$ in the expansion of $\lam (t)$ about $t_0$. We shall justify
neglecting these higher order terms shortly.

Eq. (\ref{ham2}) describes a linear quench of the system with
$\tau_{\rm eff}(\al) = \tau/(\al g^{(\al-1)/\al})$. Hence one can
use the well-known results of Landau-Zener dynamics \cite{lz1} to
write an expression for the defect density,
\bea n = \int_{{\rm BZ}} \frac{d^dk}{(2\pi)^d} p_{\vk} = \int_{{\rm BZ}}
\frac{d^d k}{(2\pi)^d} \exp[ -\pi |\Delta(\vk)|^2 \tau_{\rm eff}(\al)].
\label{efq1} \eea
For a slow quench, the contribution to $n$ comes from $\vk$ near $\vk_0$; hence
\bea n ~\sim ~\tau_{\rm eff}(\al)^{-d/2} ~=~ \left(\al g^{(\al-1)/\al}/\tau
\right)^{d/2}. \label{defect4} \eea
Note that for the special case $\al=1$, we get back the familiar result $n
\sim \tau^{-d/2}$, and the dependence of $n$ on the non-universal constant
$g$ vanishes. Also, since the quench is effectively linear, we can use the
results of Ref. \onlinecite{anatoly1} to find the scaling of the defect
density when the critical point at $t=t_0$ is characterized by arbitrary
$\nu$ and $z$,
\bea n ~\sim ~\left(\al g^{(\al-1)/\al}/\tau\right)^{\nu d/(z\nu +1)}.
\label{defect5} \eea

Next we justify neglecting the higher order terms $R_n$. We note that
significant contributions to $n$ come at times $t$ when the
instantaneous energy levels of $H'(t)$ in Eq. (\ref{ham2}) for a given
$\vk$ are close to each other, i.e., $(t-t_0)/\tau \sim \Delta(\vk)$. Also,
for a slow quench, the contribution to the defect density is substantial
only when $p_{\vk}$ is significant, namely, when $|\Delta(\vk)|^2 \sim
1/\tau_{\rm eff}(\al)$. Using these arguments, we see that
\bea R_n/R_{n-1} &=& (\al-n+1)g^{-1/\al}(t-t_0)/(n\tau) \non \\
& \sim & (\al-n+1)/(n \sqrt{\tau}). \label{negterm2} \eea
Thus we find that all higher order terms $R_{n>1}$, which were neglected in
arriving at Eq. (\ref{defect4}), are unimportant in the limit of slow quench
(large $\tau$).

The scaling relations for the defect density $n$ given by Eqs.
(\ref{defect3}) and (\ref{defect5}) represent the central results of
this section. For such power law quenches, unlike their linear
counterpart, $n$ depends crucially on whether or not the quench term
vanishes at the critical point. For quenches which do not vanish at
the critical point, $n$ scales with the same exponent as that of a
linear quench, but is characterized by a modified non-universal
effective rate $\tau_{\rm eff}(\al)$. If, however, the quench term
vanishes at the critical point, we find that $n$ scales with
a novel $\al$-dependent exponent $ \al d \nu/(\al z \nu +1)$. For
$\al=1$, $\tau_{\rm eff}(\al) = \tau$ and $\al d \nu/(\al z \nu + 1)
= d \nu/(z\nu +1)$; hence both Eqs. (\ref{defect3}) and
(\ref{defect5}) reproduce the well-known defect production law for
linear quenches as a special case \cite{anatoly1}. We note that the
scaling of $n$ will show a cross-over between the expressions given
in Eqs. (\ref{defect3}) and (\ref{defect5}) near some value of $\tau
= \tau_0$ which can be found by equating these two expressions; this
yields $\tau_0 \sim |b (\vk_0)|^{- z \nu - 1/\al}$. For $\al
> 1$, the scaling law will thus be given by Eq. (\ref{defect3})
(Eq. (\ref{defect5})) for $\tau \ll(\gg) \tau_0$. We also note here
that the results of this section assumes that the system passes from
one gapped phase to another through a critical point and do not
apply to quenches which take a system along a critical line
\cite{pell1,sen2}. We shall deal with this case in Sec. \ref{hamil3}.

\subsection{Quench dynamics through a multicritical point} \label{hamil2}

In this section, we will consider the effect of a nonlinear quench in a
system of the form given in Eq. (\ref{ham1}), except that we now take
\beq b(\vk) \sim |\vk - \vk_0|^{z_1}, ~~~{\rm and}~~~ \Delta(\vk) \sim
|\vk - \vk_0|^{z_2}, \label{general} \eeq
so that the system passes through the critical point at $t=0$.
This will be a generalization of the discussion in the first part of Sec.
\ref{hamil1} where we had $z_1 > z_2$ with $z_2 = 1$. We will see below that
a separate analysis is required if $z_2 > z_1$. As discussed recently in Ref.
\onlinecite{sen4}, such a condition arises at the multicritical point of
a one-dimensional spin-1/2 $XY$ model in a transverse field; in that
model, we find that $z_1 = 2$ and $z_2 = 3$.

We begin our analysis by comparing the diagonal and off-diagonal
terms in Eq. (\ref{ham1}). From general considerations, it is clear
that defects are mainly produced when both $|t/\tau|^\al {\rm
sign}(t) + b(\vk)$ and $|\Delta(\vk)|$ are of order 1 or less since
this is when the instantaneous energy levels given by Eq.\ \ref{en1}
are close to each other. We now consider the forms of $b(\vk)$ and
$\Delta(\vk)$ given in Eq. (\ref{general}). Two possibilities arise
in the limit $\tau \to \infty$ and $|\vk - \vk_0| \to 0$.

\noi (i) If $z_1 > z_2$, then $|\Delta(\vk)|$ being of order 1 or
less implies that $b(\vk)\ll |\Delta(\vk)|$, namely, $b(\vk)\ll 1$.
In this case, we can ignore the term $b(\vk)$ in Eq. (\ref{ham1}).
This is equivalent to saying that the first argument of the scaling
function $f$ in Eq. (\ref{pexp}) can be set equal to zero. Following
arguments similar to those leading up to Eq. (\ref{sca1}), we then
see that the defect density scales as \beq n ~\sim~ \tau^{-d \al
/[z_2 (\al + 1)]} \label{sca2} \eeq which is independent of the
value of $z_1$.

\noi (ii) If $z_2 > z_1$, then $|\Delta(\vk)|$ being of order 1 or
less implies that $b(\vk)\gg |\Delta(\vk)|$, namely, $b(\vk)\gg 1$.
Thus $b(\vk)$ always remains finite as we approach the critical
point and cannot in general be neglected. In order to have
$|t/\tau|^\al {\rm sign}(t) + b(\vk)$ of order 1 or less, we must
therefore have $t \gg 1$. Let us define a time $t_0$ as
$|t_0/\tau|^\al = - {\rm sign(t_0)} b(\vk) = - {\rm sign(t_0)} \eta
|\vk - \vk_0|^{z_1}$, where $\eta$ is an arbitrary non-universal
constant. Thus \beq |t_0| ~=~ |\eta|^{1/\alpha} \tau |\vk -
\vk_0|^{z_1/\al}. \label{t0} \eeq In a spirit similar to Eq.
(\ref{ham2}), we now linearize the function $|t/\tau|^\al {\rm
sign}(t) + b( \vk)$ near $t=t_0$, as $(|t/\tau|^\al {\rm sign(t)}-
|t_0/\tau|^\al{\rm sign(t_0)}) = (t-t_0) \al |t_0/\tau|^{\al
-1}/\tau$ which, using Eq. (\ref{t0}), is equal to $(t-t_0)/
\tau_{\rm eff}(|\vk|;\alpha)$ where \bea \tau_{\rm
eff}(|\vk|;\alpha) \equiv \tau_{\rm eff} &=& \tau
|\eta|^{(\alpha-1)/\alpha} |\vk - \vk_0|^{-z_1 (\al -
1)/\al}/\alpha.\nonumber\\ \label{efftau1} \eea The effective
linearized Hamiltonian can be written as \bea H_{\rm eff}&=&
\left(\tau_3 (t-t_0)/\tau_{\rm eff} + |\vk -\vk_0|^{z_2} \tau_1
\right) \eea and describes a linear quench with $\tau$ replaced by
$\tau_{\rm eff}(|\vk|;\al)$. The corresponding defect density is
therefore given by the Landau-Zener expression in Eq. (\ref{efq1}).
We find that \beq p_{\vk} ~\sim ~\exp ~[- ~\pi ~ \tau ~ |\vk
-\vk_0|^{[2z_2 - z_1 (1-1/\al)]} ~ |\eta|^{(\alpha-1)/\alpha}/ \al],
\eeq and \beq n ~\sim~ \tau^{-d \al /[2z_2 \al + z_1 (1 - \al)]}.
\label{sca3} \eeq Note that the defect density obtained in Eq.
(\ref{sca3}) scales with an exponent which is independent of the
non-universal coefficient $\eta$.

To generalize these results for models with arbitrary $z_1<z_2$ and
$\nu$, we notice that such models can be described by an effective
Hamiltonian $H_{\rm eff}(\lambda(t))$, where $\lambda(t) =
(t-t_0)/\tau_{\rm eff} (|\vk|;\alpha)$ and $\tau_{\rm eff}(|\vk|
;\alpha)$ is given by Eq. (\ref{efftau1}). This effective
Hamiltonian therefore describes a linear quench with a different
$\tau_{\rm eff}$ for each $\vk$ mode and with effective dynamical
critical exponent $z_2$ and correlation length exponent $\nu$. Thus
using the arguments of Ref. \onlinecite{anatoly1}, we get
\bea n &\simeq& \int \frac{d^d k}{(2 \pi)^d} \Big| \int_{-\infty}^\infty d
\lam \la \vk|\frac{d}{d \lam} |0 \ra e^{i \tau_{\rm eff}
(|\vk|;\alpha) \int^\lam d \lam' \de E_{\vk} (\lam')} \Big|^2, \non \\
& & \label{taueff2} \eea where $\de E_{\vk} (\lam) \simeq |\lam|^{
z_2 \nu} F'(|\lam|^{z_2 \nu}/|\vk|^{z_2})$, and $F'(x) \sim 1/x$ for
large $x$. Further, one has $\la \vk|\frac{d}{d \Delta} |0\ra =
|\vk|^{-z_2} G(\Delta /|\vk|^{z_2})$ near a critical point, where
$G(0)$ is a constant. Using these relations, one obtains $n \simeq
\int d^dk \Big| \int d\lambda' \lambda^{' z_2\nu-1} G(\lambda') \exp
\Big( \frac{i \tau}{\alpha} |\vk|^{[\alpha(z_2\nu+1) + z_1 \nu (1 -
\alpha)]/\alpha \nu}$
\\ $\int^{\lambda'} d\lambda'' \lambda'' F(\lambda'' ) \Big) \Big|^2$,
where $\lambda'=\lambda/|\vk|^{1/ \alpha \nu}$ and we have set $\eta=1$
without any loss of generality. Then scaling $|\vk| \to |\vk|
\tau^{\alpha \nu/\left[\alpha (z_2\nu+1) + z_1 \nu (1 -
\alpha)\right]}$, one finally gets \bea n ~\sim ~\tau^{-d \alpha
\nu/\left[ \alpha(z_2\nu+1) + z_1 \nu (1 - \alpha) \right]},
\label{scaa4} \eea which reduces to Eq. (\ref{sca3}) for $z_2 \nu
=1$. Note that for Eq. (\ref{sca2}), a generalization to models with
arbitrary $z_2 \nu$ is straightforward, and is given by Eq.
(\ref{defect3}) with $z$ replaced by $z_2$.

Eqs. (\ref{sca2}), (\ref{sca3}) and (\ref{scaa4}) are the main
results of this section. These results generalize those in Sec.\
\ref{hamil1} to defect production for quenches through arbitrary
multicritical points. Note that for $z_1 = z_2$, Eq. (\ref{sca2})
and Eq. (\ref{sca3}) agree for any value of $\al$, giving $n \sim
\tau^{-d \al /[z_2 (\al + 1)]}$. Further, for the case of linear
quenching, $\al =1$, these equations agree for any value of $z_1$
and $z_2$, giving $n \sim \tau^{-d /(2z_2)}~$ which has been
recently obtained in Ref. \onlinecite{sen4}.

\subsection{Quench dynamics along a gapless line}
\label{hamil3}

Recently quench dynamics in a one-dimensional $XY$ model in the presence
of a spatially modulated transverse magnetic field has been studied in
Ref. \onlinecite{sen3}. Such a model is described by the Hamiltonian
\bea H &=& -~\frac{1}{2} ~\sum_{j} ~\Big[ J \left(\sigma_j^x \sigma_{j+1}^x +
\sigma_j^y \sigma_{j+1}^y \right) \non \\
&& + \gamma \left(\sigma_j^x \sigma_{j+1}^x -\sigma_j^y \sigma_{j+1}^y \right)
- \left(h-(-1)^j \de\right) \sigma_j^z \Big], \label{xyham} \eea
where $J$ and $\gamma$ are respectively the strength of and the anisotropy
in the nearest neighbor spin-spin interactions, $\sigma^a$ ($a=x,y,z$)
denote the Pauli matrices, and $h$ and $\de$ denote the uniform and
alternating components of the magnetic fields respectively. The phase diagram
of this model is discussed in detail in Ref. \onlinecite{sen3}. It was pointed
out that quenching the anisotropy parameter $\gamma(t)=\gamma_0 t/\tau$
linearly while sitting at the paramagnetic phase determined by the
condition $h^2=\de^2+J^2$, leads to a time evolution of the
system along a gapless line. It was also shown, via mapping this spin model
to a system of Majorana fermions by a Jordan-Wigner transformation,
that the evolution of the model described by Eq. (\ref{xyham}) can be
represented by an effective Hamiltonian given by \cite{sen3}
\bea H_{\rm eff}(k;t) &=& \sum_k \psi_k^{\da} \left({\tilde
\gamma}(t)k \tau_3 + {\tilde J} k^2 \tau_1 \right) \psi_k , \eea
where $\psi_k = (c_{1k},c_{2k})$ is the usual two component fermionic field,
${\tilde \gamma}(t) = \gamma(t)J/\sqrt{\de^2+J^2}$ and ${\tilde J} =
J^2/\sqrt{\de^2+J^2}$. The quench dynamics of this model was studied in Ref.
\onlinecite{sen3} for the linear quench $\gamma(t)=\gamma_0t/\tau$ using the
Landau-Zener formalism. It was found that the defect density scales as
\bea n ~\sim ~\tau^{-1/3}. \eea
Note that since for this model $z=\nu=1$, the naive expectation
according to the analysis of Sec. \ref{hamil1} is to have $n \sim
1/\sqrt{\tau}$. This result therefore clearly points out the
necessity of extending the analysis of Sec. \ref{hamil1} for
quenches along gapless lines in parameter space. In what follows,
we shall only restrict ourselves to quenches where the gap vanishes
at the same momentum value $k_0$; $k_0=0$ for the present case.

We start with a generic $d$-dimensional model described by a Hamiltonian
\bea H_{\rm eff}(k;t) &=& \sum_k \psi_k^{\da} \left(\lam (t)
|\vk|^a \tau_3 + \Delta_0 |\vk|^b \tau_1 \right) \psi_k , \eea
where $\lam (t) = \lam_0 |t/\tau|^{\al} {\rm sign}(t)$ is the
quench parameter, $a$ and $b$ are arbitrary exponents, and we have
taken $\vk_0=0$ for clarity. Note that $d=a=1$ and $b=2$ corresponds
to the $XY$ model studied in Ref. \onlinecite{sen3}, while $a=0$ and
$d=b=1$ corresponds to the one-dimensional Kitaev model studied in
Ref. \onlinecite{sen2}. For $a\ne0$, the system passes along a gapless
line during the quench. We study the time evolution of the model in a manner
similar to that described in Sec. \ref{hamil1}. After some straightforward
algebra, one obtains the equation for the evolution of $c'_{1\vk}(t)=
c_{1\vk}(t) \exp (i \int^t dt^{'}|t^{'}/\tau|^{\al} {\rm sign} (t^{'})
|\vk|^{a})$ as
\beq \ddot{c}_{1\vk}^{'} - 2i|t/\tau|^{\al}{\rm sign}(t) |\vk|^a ~
\dot{c}_{1\vk}^{'} + \Delta_0^2 ~|\vk |^{2b} ~c_{1\vk}^{'}= 0.
\label{c1eqn} \eeq
Next we define $\tau_{\vk} = \tau/k^{a/\al}$ and scale $t \to t
\tau_{\vk}^{\al/(\al+1)}$ in Eq. (\ref{c1eqn}) to obtain
\bea \ddot{c'}_{1\vk} &-& 2i|t|^\al {\rm sign} (t) ~\dot{c'}_{1\vk} \non \\
&+& |\Delta|^2_0 ~|\vk|^{2b-
\frac{2a}{\al+1}}\tau^{\frac{2\al}{\al+1}} ~ c'_{1\vk}=0.
\label{c1eqn2} \eea {}From Eq. (\ref{c1eqn2}), we find that the
probability of defect formation for a given momentum $\vk$ must be
given by \bea p_{\vk} &=& \lim_{t\to \infty} |c'_{1\vk}(t)|^2 = f[
\Delta_0^2 | \vk |^{2b-2a/(\al +1)} \tau^{2\al/(\al+1)} ], \non \\
\eea where $f[\infty]=0$. The defect density therefore becomes \beq
n ~\sim ~\int_{BZ}\frac{d^dk}{(2\pi)^d} ~f(\Delta_0^2 \tau^{\frac{2
\al}{\al+1}}|\vk|^{2b-\frac{2a}{\al+1}}). \eeq Using the same logic
as outlined in Sec. \ref{hamil1}, we scale $|\vk| \to
\tau^{\frac{\al}{b(\al+1)-a}} |\vk| $ and get \beq n ~\sim
~\tau^{-\frac{d\al}{b(\al+1)-a}} . \label{sca4} \eeq This result
generates the scaling of the defect density derived in Ref.
\onlinecite{sen3} ($n\sim\tau^{-d/(2b-a)}$) for the special case
$\al=1$, and that of the one-dimensional Kitaev model studied in
Ref. \onlinecite{sen2} for $\al =b=z=1$ and $a=0$.

Finally, we generalize the result in Eq. (\ref{sca4})
to systems where the energy difference between the ground and excited
states vanishes along the gapless line as $\Delta=\lam^{\al
c}|\vk|^a$. Note that for the quenches treated in Sec. \ref{hamil1},
$c=z\nu$. Here, however, since the quench takes place along a gapless line,
$c$ need not have the same interpretation and can be system specific. Exactly
at the quantum critical point $\lam=\lam_c$, the energy gap vanishes as
$\Delta\sim |\vk|^b$. Thus $b$ is to be interpreted as the dynamical scaling
exponent $z$ in the present case. Then using the same scaling argument as in
Sec. \ref{hamil1}, we can express the defect density $n$ using Eq.
(\ref{defect2}). However, in the present case the energy gap scales as
\bea E_k(\lam) -E_0(\lam) &=& \Delta F(\Delta/|k|^b), \non \\
\la \vk |\frac{d}{d\Delta}|0\ra &=& 1/|k|^b G(\Delta/|k|^b) , \label{ss1} \eea
where $F(x)=1/x$ for large $x$, and $G(0)$ is a constant. Using Eq.
(\ref{ss1}), we see that
\bea E_k(\lam) -E_0(\lam)&=&\lam^{\al c}|\vk|^a F(\lam^{\al c}|\vk|^a/|\vk|^b),
\non \\
\la \vk |\frac{d}{d\lam}|0\ra &=& \la \vk |\frac{d}{d\Delta}|0\ra
\frac{d\Delta}{d\lam} \non \\
& \simeq & \frac{\lam^{\al c-1}}{|\vk|^{(b-a)}} ~G\left(\frac{\lam^{\al c}
|\vk|^a}{|\vk|^b}\right). \eea
Substituting these in Eq. (\ref{defect2}), the defect density produced in this
system is found to be
\bea n &\sim& \int \frac{d^dk}{(2\pi)^d} ~\left|\int_{-\infty}^{\infty}
\frac{\lam^{\al c-1}}{|\vk|^{(b-a)}} G\left(\frac{\lam^{\al c}} {|\vk|^{b-a}}
\right) \right. \non \\
& & ~~~~~~~~~~~\left. e^{i\tau\int^{\lam}d\lam^{'}\lam^{'\al c}|\vk|^a
F(\lam{'}^{\al c} |\vk|^a/|\vk|^b)}\right|. \eea
Defining new variables $\xi=\lam |\vk|^{(a-b)/\al c}$ and
$|\vk^{'}|=|\vk|\tau^{\al c/[(b-a)+b\al c]}$, we get
\beq n ~\sim ~\tau^{-d\al c /[(b-a)+b\al c]} . \label{defectfinal} \eeq

Eq. (\ref{defectfinal}) is one of the central results of this work,
and it generates all the previous scaling laws for both linear and
nonlinear quenches through critical lines and points (but not through
multicritical points) as special cases. For $a=0$, $b=z$ and
$c=z\nu$, we recover the scaling law Eq. (\ref{defect3}) for a
nonlinear quench, whereas for $\al =c=1$, we obtain the scaling law
derived for a linear quench in Ref. \onlinecite{sen3}.

\section{Defect correlation functions}
\label{c1}

For the purpose of computation of defect correlation functions, we
are going to restrict ourselves to the class of $d$-dimensional
models given by $H(t)$ in Eq. (\ref{ham1}). As mentioned before,
many standard spin models in one and two dimensions can be mapped,
via standard Jordan-Wigner transformations \cite{sachdev1}, to such
fermionic models described by $H(t)$. Let us denote the ground and
the excited states of $H(t)$ before the quench (at $t = -\infty$) by
$|0\ra_{\vk}$ and $|1\ra_{\vk}$ respectively for a given value of
$\vk$. Then the state of the system after the quench (at $t =
\infty$) is given by \cite{sen2} \bea |\phi\ra_{\vk} ~=~
\sqrt{p_{\vk}} ~|0 \ra_{\vk} ~+~ \sqrt{1-p_{\vk}} ~|1 \ra_{\vk} .
\eea Using this, one can compute the defect correlation functions
for these models. These correlation functions are of two types. They
can either vanish at the origin, as in the case of the
two-dimensional extended Kitaev model \cite{sen2}, or can be written
as \cite{dziar1,sen2} \bea \la O_{\vcr}\ra = -\de_{{\vcr},0}+ C
\int_0^{2\pi} d^d k f [ |\Delta (\vk) |^2\tau^{\frac{2\al}{\al+1}} ]
g (\vk \cdot \vcr), \label{corrgen1} \eea where $O_{\vcr}= i
\psi_{\vn} \psi_{\vn + \vcr}$, $\psi_{\vn}$ denotes the field
operators for Majorana fermions, $g(\vk \cdot \vcr)$ is a system
specific function independent of $\tau$, $C$ denotes a system and
dimension specific constant which will be unimportant for subsequent
discussions, and we have used Eq. (\ref{pexp}) to obtain the value
of $p_{\vk}$. Since for a slow quench, $p_{\vk}$ is appreciable only
near $\vk=\vk_0$, we expand $\Delta(\vk)$ about $\vk_0$, scale the
momentum components $ k'_i = (k-k_0)_i \tau^{\al/(\al+1)}$, and
extend the range of integration to $\infty$ to get \bea \la
O_{\vcr}\ra = - \de_{{\vcr},0} + \frac{C}{\tau^{d\al/(\al+1)}}
\int_0^{\infty}d^d k' f[|\vk^{'}|^2 ] g (\vk^{'} \cdot \vcr^{'}),
\non \\ \label{corrgen2} \eea where ${\vcr}^{'}_i=
{\vcr}_i/\tau^{\al/(\al+1)}$. Thus we find that quite generally, for
the class of models whose defect correlation functions do not vanish
at the origin, \bea \ln\left(1+\la O_{{\vcr}=0} \ra\right) ~=~
\ln(C') -\frac{d\al}{\al+1} \ln(\tau) , \eea i.e., the logarithm of
the deviation of the amplitude of these correlation functions at the
origin from $-1$ is a linear function of $\ln(\tau)$ with a slope of
$-d \al/(\al+1)$.

\begin{figure}
\includegraphics*[width=\linewidth]{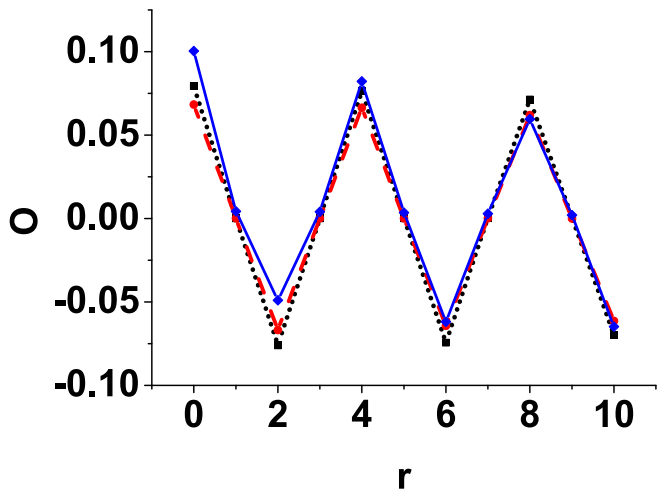}
\caption{Plot of $\la O_r \ra$ vs $r$ for $\al=2$ (black dot-dashed line),
$\al=3$ (red dashed line), $\al=4$ (blue solid line), and $\tau=20$.}
\label{correl1d} \end{figure}

We now compute the correlation function for a specific model, namely, the
one-dimensional Kitaev model [\onlinecite{sen2,ks3,kit1}] which has the
Hamiltonian
\beq H ~=~ \sum_{i\in \rm{even}} ~\left( J_1 S_i^x S_{i+1}^x ~+~ J_2 S_i^y
S_{i-1}^y \right), \label{h1} \eeq
where $J_1$ and $J_2$ denote the nearest neighbor interaction strengths, and
$S_i$ denotes the spin at site $i$. Using the standard Jordan-Wigner
transformation, this Hamiltonian can be mapped on to a free fermionic
Hamiltonian [\onlinecite{sen2,ks3,kit1}]
\bea H &=& \sum_{\vk} ~\psi^{\da}_k ~H_k ~\psi_k, \hspace{0.3cm} \rm{where}
\non \\
H_k &=& -2 ~(J_-\sin(k) ~\tau_3+J_+\cos(k) ~\tau_2) . \label{h1'} \eea
Here $J_{\pm}=J_1\pm J_2$, and $\psi_k=(c_1(k),c_2(k))$ are the fermionic
fields. The Hamiltonian is changed in time by varying the parameter
$J_-$ keeping $J_+$ fixed. The defect correlation function for this
model is given by [\onlinecite{sen2}]
\beq \la O_r\ra ~=~ -~\de_{r,0} ~+~ \frac{2}{\pi} ~\int_0^{\pi} ~dk ~p_k~
\cos(kr). \label{correl} \eeq
Thus we find that the defect correlation functions have the same form
as in Eq. (\ref{corrgen1}) with $C=2/\pi$ and $g=\cos(k r)$. A plot of
the correlation function as a function of $r$, sans the $\de$-function peak
at the origin, is shown in Fig. \ref{correl1d} for
$\tau=20$ and several representative values of $\al$.

\section{Numerical evaluation of defect densities}
\label{num1}

In this section, we provide numerical studies of the one-dimensional
Ising and Kitaev models to supplement our analytical results. First
we consider the one-dimensional Ising model in a transverse field
described by \beq H_{\rm Ising} ~=~ - ~J ~(\sum_i ~S_i^z S_{i+1}^z
~+~ g ~\sum_i ~S_i^x), \eeq where $J$ is the strength of the nearest
neighbor interaction, and $g=h/J$ is the dimensionless transverse
field. In what follows, we shall quench the transverse field as
$g(t)= |t/\tau|^{\al} {\rm sign}(t)$ and compute the density of the
resultant defects.

\begin{figure}
\rotatebox{0}{\includegraphics*[width=\linewidth]{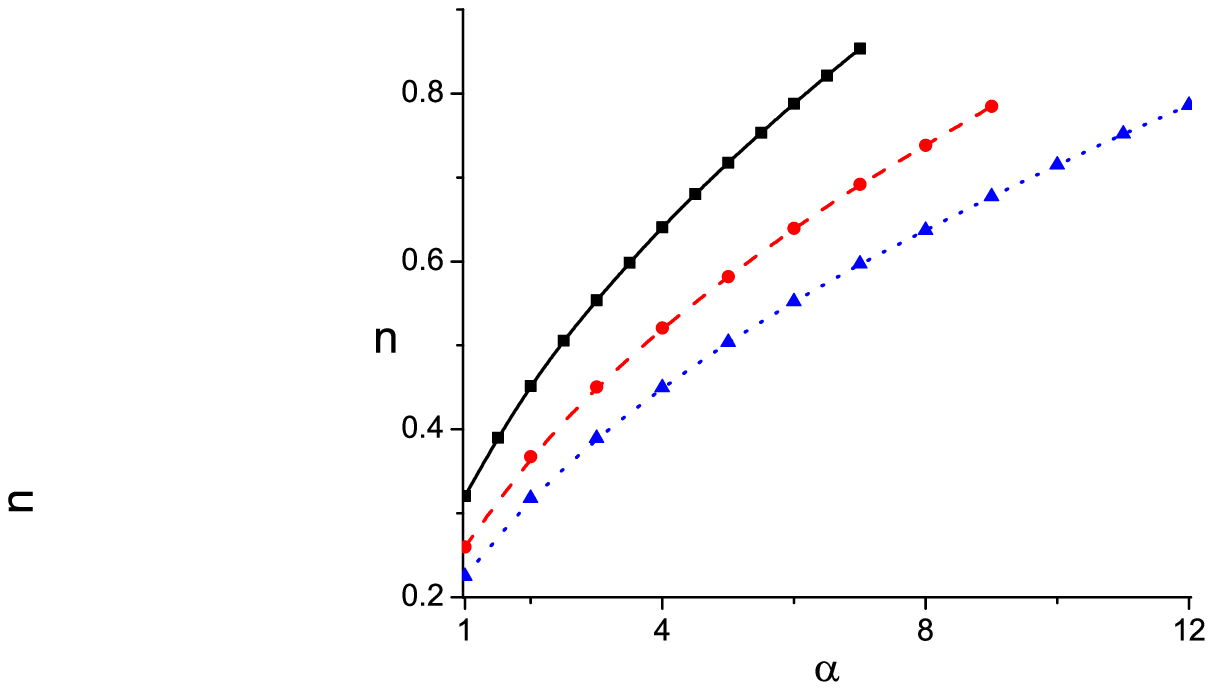}}
\caption{Variation of the defect density $n$ with the quench exponent $\al$
for representative values of $\tau=10$ (black solid line), $\tau=15$ (red
dashed line) and $\tau=20$ (blue dotted line). A polynomial fit of the form
$n = a \al^b$ yields exponents which are very close to the theoretical result
$1/2$ for all values of $\tau$ (see text for details).} \label{fig1}
\end{figure}

We begin by mapping $H_{\rm Ising}$ to a system of free fermions via a
standard Jordan-Wigner transformation \cite{sachdev1}
\beq H^{'} ~=~ - ~J\sum_k ~\psi_k^{\da}((g-\cos(k))\tau_3+\sin(k)\tau_1)
\psi_k. \eeq
If the external magnetic field $g$ is varied
with time as $g(t)=g_0|t/\tau|^{\al} {\rm sign}(t)$, then
the system will go through two quantum critical points at $g= 1$ and $-1$.
The energy gap vanishes at these quantum critical points at $k=k_0=0$ and
$\pi$. As a result, defects are produced in non-adiabatic regions near these
points. For this model, the quantum critical point is at $t=t_0\ne 0$ and
$z=\nu=1$. Hence, $\tau_{{\rm{eff}}}= \tau/\al$ for both the quantum critical
points. From Eq. (\ref{defect5}), therefore, we expect the defect density
produced in this system to be given by $n\sim(\tau/\al)^{-1/2}$.

To verify this expectation, we numerically solve the Schr\"odinger
equation $i\partial_t \psi_k = H_k (t) \psi_k$ and obtain the
probability $p_k$ for the system to be in the excited state.
Finally, integrating over all $k$ within the Brillouin zone, we
obtain the defect density $n$ for different values of $\al> 1$ with
fixed $\tau$. The plot of $n$ as a function of $\al$ for
$\tau=10,~15$ and $20$ is shown in Fig. \ref{fig1}. A fit to these
curves gives the values of the exponents to be $0.506 \pm 0.006$,
$0.504 \pm 0.004$ and $0.505 \pm 0.002$ for $\tau=10,~15$ and $20$
respectively which are remarkably close to the theoretical value
$1/2$. The systematic positive deviation of the exponents from the
theoretical value $1/2$ comes from the contribution of the higher
order terms neglected in the derivation of Eqs. (\ref{defect4}) and
(\ref{defect5}). We note that the region of validity of our linear
expansion, as can be seen from Fig. \ref{fig1}, grows with $\tau$
which is in accordance with the result in Eq. (\ref{negterm2}).

\begin{figure}
\rotatebox{0}{\includegraphics*[width=\linewidth]{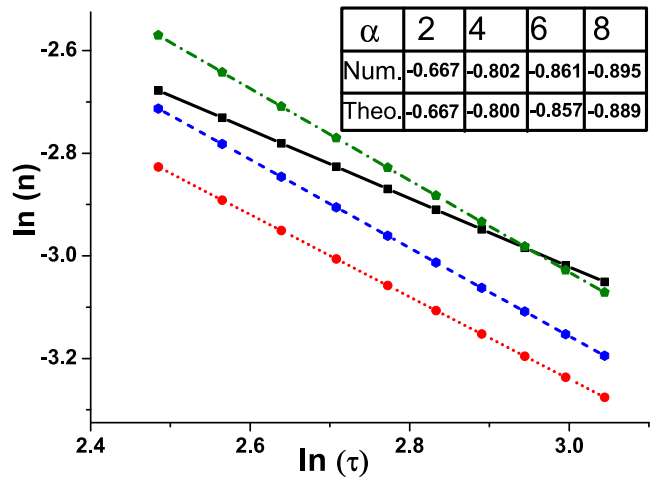}}
\caption{Plot of $\ln(n)$ vs $\ln(\tau)$ for the one-dimensional Kitaev model
for $\al=2$ (black solid line), $\al=4$ (red dotted line), $\al=6$ (blue
dashed line) and $\al=8$ (green dash-dotted line). The slopes of these lines
agree reasonably with the predicted theoretical values $-\al/(\al+1)$ as
shown in the table.} \label{kitaev} \end{figure}

Next, we consider the one-dimensional Kitaev model which is governed by
the Hamiltonian in Eq. (\ref{h1}). As mentioned in Sec. \ref{correl},
such a model can also be mapped on to the free fermionic Hamiltonian
given by Eq. (\ref{h1'}). This system passes through the quantum
critical point at $J_-=0 $ for $k=\pi/2$ when $J_-(t)= J_- |t/\tau|^\al
{\rm sign}(t)$ is varied nonlinearly with time. Here the quantum critical
point is at $t=0$. Thus from Eq. (\ref{defect3}) we expect the
defect density $n\sim\tau^{-\al/(\al+1)}$ since $\nu=z=1$ for this
system. To check this prediction, we numerically solve the
Schr\"odinger equation $i \partial_t \psi(k) = H' (k;t) \psi(k,t)$
and compute the defect density $n = \int_0^{\pi} dk/\pi \, p_k$ as
a function of the quench rate $\tau$ for different $\al$ with fixed $J_+
/J=1$. A plot of $\ln (n)$ vs $\ln (\tau)$ for
different values of $\al$ is shown in Fig. \ref{kitaev}. The slope of
these lines, as can be seen from Fig. \ref{kitaev}, changes from
$-0.67$ towards $-1$ as $\al$ increases from $2$ towards larger values.
This behavior is consistent with the prediction of Eq. (\ref{defect3}).
The slopes of these lines also show excellent agreement with
Eq. (\ref{defect3}) as shown in the inset of Fig. \ref{kitaev}.

\begin{figure}
\rotatebox{0}{\includegraphics*[width=\linewidth]{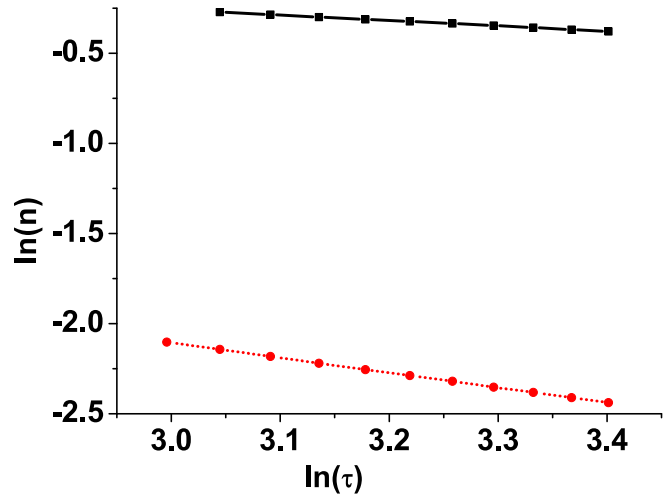}}
\caption{Plots of $\ln(n)$ vs $\ln(\tau)$ for models with $z_1 = 2$ and $z_2
= 1$ (red dotted lower line) and $z_1 = 1$ and $z_2 = 2$ (black solid upper
line), for $d=1$ and $\alpha = 4$. The slopes of the lower and upper lines are
$-0.828$ and $-0.301$ which compare reasonably with the predicted theoretical
values of $-4/5 = - 0.8$ and $-4/13 = - 0.308$ respectively.} \label{z1z2}
\end{figure}

Finally, we illustrate the expressions in Eqs. (\ref{sca2}) and
(\ref{sca3}) by taking two one-dimensional models governed by Eqs.
(\ref{ham1}) and (\ref{general}) with $z_1 = 2$, $z_2 = 1$ and $z_1
= 1$, $z_2 = 2$ respectively. Setting $\al =4$, we numerically carry
out the time evolutions for different values of the momentum $k$ and
then integrate to compute the defect density as a function of
$\tau$. The results are shown in Fig.\ \ref{z1z2}; reasonable
agreement is obtained with the theoretical values of the exponents
given in Eqs. (\ref{sca2}) and (\ref{sca3}).

\section{Experiments}
\label{expts}

The generality of our results allows for their verification in
several realizable experimental systems. We note that all our results
have been obtained at zero temperature with the assumption that the
system does not relax significantly during the quench process and till
the measurement of the defect density has been performed. This might
seem too restrictive. However, we would like to point out that
systems of ultracold atoms in optical or magnetic traps and/or optical
lattices can easily satisfy the required criteria since they have a
very long relaxation time which often gets close to the system
lifetime \cite{bloch1}. We list some possible experiments briefly
here. First, there has been a concrete proposal for the realization
of the Kitaev model using an optical lattice\cite{duan}. In such a
realization, all the couplings can be independently tuned using
separate microwave radiations. In the proposed experiment, one needs
to keep $J_3=0$ and vary $J_{1(2)} = J(1\pm |t/\tau|^{\al} {\rm
sign}(t))/2$, so that $J_+$ remains constant while $J_-$ varies in
time. The variation of the defect density, which in the experimental
set-up would correspond to the bosons being in the wrong spin state,
would then show the theoretically predicted power law behavior in Eq.
(\ref{defect3}). Secondly, a similar quench experiment can be
carried out with spin-1 bosons in a magnetic field described by an
effective Hamiltonian $H_{\rm eff} = c_2 n_0 \la {\bf S} \ra^2 + c_1
B^2 \la S_z^2 \ra~$ \cite{sk1}, where $c_2 < 0$ and $n_0$ is the boson
density. Such a system undergoes a quantum phase transition from a
ferromagnetic state to a polar condensate at $B^{\ast} =
\sqrt{|c_2|n_0/c_1}$. A quench of the magnetic field $B^2=B_0^2
|t/\tau|^{\al}$ would thus lead to a scaling of the defect density with an
effective rate $\tau_{\rm eff} (\al)= \tau/(\al g^{(\al-1)/\al})$,
where $g=|c_2|n_0 /c_1$. A measurement of the dependence of the
defect density $n$ on $\al$ should therefore serve as a test of
the prediction in Eq. (\ref{defect5}). Finally, spin gap dimer compounds
such as ${\rm Ba Cu Si_2 O_6}$ are known to undergo a
singlet-triplet quantum phase transition at $B_c \simeq 23.5 $T
which is known to be very well described by the mean-field exponents
$z=2$ and $\nu=2/3~$ \cite{sebas1}. Thus a nonlinear quench of the
magnetic field through its critical value $B=B_c + B_0
|t/\tau|^{\al}{\rm sign}(t)$ should lead to a scaling of the defects $n
\sim \tau^{-6\al/(4\al+3)}$ in $d=3$. In the experiment, the defect
density would correspond to residual singlets in the final state
which can be computed by measuring the total magnetization of the
system immediately after the quench. We note that for these dimer
systems, it will be necessary to take special care to achieve the
criterion of long relaxation time mentioned earlier.

\section{Conclusions}
\label{conclusion}

In conclusion, we have studied defect production in quantum critical
systems for an arbitrary nonlinear power law quench. We have shown
that the defect production rate depends crucially on whether the
system passes from one gapped phase to another or along a critical
gapless line during the quench. We have obtained general scaling
laws for defect densities produced during the quench for both these
cases, and have verified these laws by numerical studies of
one-dimensional systems. We have also computed the defect
correlation functions for a class of $d$-dimensional models and have
discussed the scaling of the amplitude of these functions with the
quenching rate. Finally, we have discussed several experimental systems
where these results can, in principle, be tested.

The authors thank A. Dutta and A. Polkovnikov for helpful comments and
discussions. DS acknowledges financial support from DST, India under Project
No. SR/S2/CMP-27/2006.

\end{document}